\documentclass[aps,prl,twocolumn,floatfix, showpacs]{revtex4}
\usepackage{graphicx}
\usepackage{dcolumn}
\usepackage{bm}

\newcommand{\be}{\begin{equation}}
\newcommand{\ee}{\end{equation}}

\begin{document}

\title{A model for the large-scale circulation in turbulent Rayleigh-B{\'e}nard convection}
\author{Eric Brown}
%\email{ebrown@physics.ucsb.edu}
\author{Guenter Ahlers}
\affiliation{Department of Physics and iQCD, University of California, Santa Barbara, CA 93106}
\date{\today}
 
\begin{abstract}

A model for the large-scale circulation (LSC) dynamics of turbulent Rayleigh-B{\'e}nard convection is presented.  It consists of two stochastic ordinary differential equations motivated 
by the Navier-Stokes equation, one each for the strength and the azimuthal orientation of the LSC. Stochastic 
forces represent phenomenologically the action of the turbulent fluctuations on the LSC. Without adjustable 
parameters, the model yields a meandering LSC with occasional rotations, and with more rare cessations that occur a 
few times per day -- differing only by about a factor of two from experiment.  Also as in experiments, the distribution 
of LSC orientation-change is uniform for cessations and a power law for rotations.  
\end{abstract}

\pacs{47.27.-i, 05.65.+b, 47.27.te, 47.27.eb}

\maketitle

Rayleigh B{\'e}nard convection (RBC) consists of a fluid-filled container heated from below \cite{Si94}.  It is well-known that cylindrical containers with aspect ratio $\Gamma \equiv D/L \simeq 1$ ($D$ is the sample diameter and $L$ the height) have a large-scale circulation (LSC), also known as the ``mean wind" \cite{KH81, SWL89, CGHKLTWZZ89, CCL96,QT01a,SBN02,FA04,TMMS05}. The LSC consists of warm up-flowing and cool down-flowing fluid on opposite sides of the sample. Its near-vertical circulation plane has an orientation $\theta_0$ that undergoes azimuthal diffusion \cite{BA06,BA06b}. An interesting aspect is that $\theta_0$, in addition to its diffusive meandering, undergoes relatively rapid oscillations \cite{FA04} and, on somewhat longer time scales,  experiences spontaneous and erratic re-orientations through an azimuthal displacement $\Delta \theta$. \cite{CCS97, NSSD01, BNA05, BA06, BA06b,XZX06}. One mechanism for re-orientation is an azimuthal {\it rotation} of the entire structure without much change of the  flow speed. \cite{CCS97} Rotations lead to a power-law distribution of $\Delta \theta$, with small rotations more likely than large ones. \cite{BNA05, BA06} The other mechanism is a {\it cessation} of the flow in which it stops briefly and then starts up again in a random new orientation, resulting in a uniform distribution of $\Delta \theta$ \cite{BNA05,BA06}.  Aside from its fundamental interest, the phenomenon of reorientation is important for instance because it occurs in natural convection of the atmosphere \cite{DDSC00}, and because it is responsible for changes in the orientation of Earth's magnetic field when it occurs in Earth's outer core \cite{GCHR99}.  

Several models were proposed to reproduce the LSC dynamics. Those of Sreenivasan {\it et al.} \cite {SBN02} and Benzi \cite {Be05} were based on similar stochastic equations that were chosen so as to produce the desired reversal events.  The equations yielded two opposite stable flow directions with transitions between them, in qualitative agreement with experiments by Sreenivasan {\it et al.} \cite{SBN02}.   However, such {\em local} flow reversals usually are not cessations or large rotations; they are equivalent to crossings of the LSC orientation past a fixed angle and correspond mostly to small orientation changes, or ``jitter" \cite{BA06}.  Another model, by Fontenele Araujo {\it et al.} \cite{FGL05}, was based on a deterministic force balance between buoyancy and drag. It produced LSC reversals ($\Delta \theta = \pi$), but did not consider azimuthal motion and thus did not reproduce the azimuthal dynamics seen in experiments.  Finally, a deterministic model based on the Boussinesq equations with slip boundary conditions and for ellipsoidal sample geometries was developed by Resagk {\it et al.} \cite{RPTDGFL06}. It neglected dissipation and diffusion, but  added these effects phenomenologically after the model was derived. The result was a set of ordinary differential equations (ODEs) with several parameters that could be tuned to produce a LSC with various dynamical states, including oscillations and chaotic meandering; but to our knowledge the existence and statistics of rotations and cessations were not explored.

We present a model consisting of two coupled stochastic ODEs, one for the 
speed (or ``strength") of the LSC and the other for the azimuthal LSC orientation. We retained the 
physically important terms of the Navier-Stokes (NS) equation and 
took volume averages. The resulting deterministic ODE for the LSC strength represents the balance between
buoyancy and drag forces and has two fixed points; one stable and the other, 
corresponding to cessations, unstable. It differs from the equations of Refs.~\cite{SBN02} and \cite{Be05} 
in that the nonlinearity is of order 3/2 rather than of the more usual cubic order. 
The second ODE describes the azimuthal motion of the LSC which is suppressed by its angular momentum. The dynamics of the model arises from the addition of stochastic forces that represent in a phenomenological sense the action of the turbulent fluctuations that exist throughout the system interior. We determined some parameters of the model 
from independent measurements that did not involve reorientations {\it per se}, determined others from theory, and did not adjust any of them arbitrarily. An interesting 
physical aspect of the model is that the angular momentum of the LSC hinders reorientations when the flow is 
vigorous; when the flow becomes feeble, then the angular momentum is small and  the stochastic forces are  
able to cause significant orientation change.  We note that the model of Resagk {\it et al.}  \cite{RPTDGFL06} suggests that the angular momentum of the LSC also plays an important role in its oscillations. Our model does not produce LSC oscillations, but yields cessations that are only slightly more 
frequent than the experimental observation of one or two per day  \cite{BNA05, BA06}. In agreement 
with experiment  \cite{BNA05, BA06} it gives a uniform probability distribution $p(\Delta \theta)$ for 
cessations and a power-law distribution for rotations. 
	
For the LSC strength we consider the velocity component $u_{\phi}$, where $\phi$ is an angle that sweeps the plane of the LSC, and describes it without azimuthal motion.  One expects the acceleration to be due to a balance between buoyancy and drag forces. Thus we include in the NS equation for $u_{\phi}$ only the acceleration,  buoyancy, and viscous drag terms, and neglect the nonlinear term \cite{FN_NL}:

\be
\dot u_{\phi} = g\alpha (T-T_0) + \nu\nabla^2 u_{\phi} \ .
\label{eq:nav}
\ee

\noindent Here $\alpha$ is the isobaric thermal expansion coefficient, $g$ the acceleration of gravity, and $\nu$ the kinematic viscosity. 

To obtain a model in the form of an ODE that describes the flow with only a few variables, we take a global average over the field variables that retains the essential physics. To carry out the average, we consider the experimental observation \cite{BNA05, BA06, BA06b, ABN06} that  the temperature of the LSC at the side 
wall at mid-height can be written as

\be
T = T_0 + \delta\cos(\theta_0 - \theta) 
\label{eq:temp}
\ee

\noindent  where the temperature amplitude $\delta$ represents 
the strength of the LSC, and where $\theta_0$ is its azimuthal orientation. 
The buoyancy acts on the entire LSC and is proportional $\delta$.  The profile is taken to be given by Eq.~\ref{eq:temp}, and proportional to the cylindrical radius $r$.
The velocity is assumed to be linear in $r$ and a step function in $\theta$.  Note that these assumptions 
about the geometry of the flow only affect the numerical prefactors in the equations, and not the 
functional form.  The drag is assumed to occur in the viscous boundary layers, so 
$\nabla^2 u_{\phi} \approx U/\lambda^2$ ($U$ is the maximum speed near the side wall, $\lambda$ 
is the boundary-layer width), where $\lambda = (L/2)\times R_{e,i}^{-1/2}$ ($R_{e,i} \equiv UL/\nu$ is the 
instantaneous Reynolds number). The volume average requires another factor of  $6\lambda/L$ since the drag 
is mainly in the boundary layers.  These approximations result in the volume-averaged equation

\be
(2/3)\dot U = (2/3\pi)g\alpha\delta - 12 \nu U R_{e,i}^{1/2}/L^2 \ .
\label{eq:uave}
\ee

\noindent Next we make the assumption that the amplitude $\delta$ is instantaneously proportional to the 
speed $U$, since both variables are measures of the LSC strength.  To determine the proportionality, we 
find the steady-state solution

\be
(2/3\pi)g\alpha\delta  = 12\nu U R_e^{1/2}/L^2 
\label{eq:delta_u}
\ee

\noindent  of Eq.~\ref{eq:uave}. Here $R_e$ is the normal steady-state Reynolds number.  In the dynamical equation, we allow the 
drag to depend on the instantaneous value $R_{e,i}$, so the drag term instantaneously 
scales as $U^{3/2}$.  We substitute Eq.~\ref{eq:delta_u} into Eq.~\ref{eq:uave}, combine all parameters into two constants, and add a noise term that represents the turbulent fluctuations of the flow to get the Langevin equation

\be
\dot\delta = \frac{\delta}{\tau_{\delta}} - \frac{\delta^{3/2}}{ \tau_{\delta}\sqrt{\delta_0}} + f_{\delta}(t)
\label{eq:lang_delta}
\ee
with
\be
\delta_0 = 18\pi \Delta T \sigma R_e^{3/2}/R \ ; \  \tau_{\delta} = L^2/(18\nu R_e^{1/2})\ .
\label{eq:parameters}
\ee

\noindent Here the Rayleigh number is $R \equiv \alpha g \Delta T L^3/\kappa \nu$ with $\Delta T$ the applied temperature difference and $\kappa$ the thermal diffusivity, and the Prandtl number is $\sigma \equiv \nu / \kappa$.

In the absence of the noise term Eq.~\ref{eq:lang_delta} has two fixed points, one unstable at $\delta=0$ and 
one stable when $\delta = \delta_0$.  In the stochastic equation this feature  reproduces some of the 
dominating behavior of the LSC;  the LSC spends most of its time meandering near the stable fixed point at $\delta_0$, 
but occasionally it ceases when fluctuations drive it close to $\delta = 0$.  

To study the predictions of the model, we consider the example $R=1.1\times 10^{10}$ and $\sigma = 4.4$ for a sample with $L = 24.76$ cm 
\cite{BNFA05}. Measurements yielded $\delta_0 = 0.25$ K and $R_e = 3700$ \cite{ABN06}. From Eq.~\ref{eq:parameters} one has $\tau_{\delta} =  85$ s and $\delta_0 = 0.10$ K. For $\delta_0$ theory and experiment are in order of magnitude agreement, which is as much as we can expect given the approximations made in the model derivation. For numerical calculations we adopt the experimental value because it presumably is more appropriate for the physical system.
To gain information about the noise intensity,  we examined the experimental mean-square 
amplitude-change $\langle (d\delta)^2\rangle$ ($\langle...\rangle$ represents a time average) over a time 
period $dt$ as a function of $dt$. For time scales that were not too large we found 
$\langle (d\delta)^2\rangle = D_{\delta}dt$ with $D_{\delta} = 3.5\times 10^{-5}$K$^2$/s, suggesting a 
diffusive process. This method was used before \cite{BA06b} to determine the diffusivity of $\theta_0$; but in the present case the diffusive scaling holds only over intermediate time scales because $\delta$ is bounded.  With this experimental input we make the noise in the model Brownian with diffusivity $D_{\delta}$, so $f_{\delta}(t)$ is Gaussian distributed with width $\sqrt{D_{\delta}/h}$ where $h$ is the time step in the simulation.  

The frequency of 
cessations is given approximately by the Arrhenius-Kramers result for diffusion over a 
potential barrier $\Delta V$ \cite{Kr40}.  Here $V \equiv -\int \dot\delta d\delta$,  
integrated over the deterministic part of Eq.~\ref{eq:lang_delta}, and
$\Delta V = V(0) - V(\delta_0) = \delta_0^2/(10\tau_{\delta})\simeq 7.3\cdot 10^{-5}$ K$^2$/s.  When $\Delta V \gg D_{\delta}$, the rate of cessations 
$\omega$ is given by 

\be
\omega = \omega_0 \exp(-\Delta V/D_{\delta})
\label{eq:cess_rate} 
\ee

\noindent where $\omega_0 = 1/(2\sqrt{2}\pi\tau_{\delta}) \simeq 1.3\cdot 10^{-3}$ s$^{-1}$ \cite{Kr40}. This yields about 14 cessations per day, an order of magnitude more than the experimental value of one or two per day. We attribute this difference to the fact that the condition  $\Delta V \gg D_{\delta}$ is not really satisfied.

The second Langevin equation describes the azimuthal motion.   The main driving force  is the turbulent noise. We estimated that angular momentum of the LSC in the $\phi$ coordinate damps the relatively slow azimuthal 
motion much more than the viscous drag across the boundary layer near the side wall. The physical explanation 
is that a rotating body has some stability due to its angular momentum and requires a larger torque to rotate 
in an orthogonal direction than a non-rotating body.  This phenomenon is represented by the transport term 
in the NS equation. Thus, neglecting the drag term (and Earth's Coriolis force; 
see \cite{BA06b}), we have

\be
\dot u_{\theta} + (\vec u \cdot \vec\nabla)u_{\theta} = 0 \ .
\ee

\noindent  A volume average gives
$(1/3)L\ddot\theta_0 = - (2/3) U\dot\theta_0$ .
The angular momentum leads to an effective damping that is proportional to the wind 
strength $U$,
which is important for understanding the azimuthal dynamics during cessations.  Again, we convert $U$ to 
$\delta$, combine the remaining parameters to get a new constant, and add a noise term representing 
turbulent fluctuations to get

\be
\ddot\theta_0 = - \frac{\dot\theta_0\delta}{\tau_{\dot\theta}\delta_0} + f_{\dot\theta}(t)
\label{eq:lang_theta}
\ee
with $\tau_{\dot\theta} = L^2/(2\nu R_e)$ .
For $R=1.1\times 10^{10}$ and $\sigma = 4.4$ the model predicts  $\tau_{\dot\theta} \simeq 13$ s.   The turbulent noise 
in this coordinate is also found to be Brownian, with diffusivity $D_{\dot\theta} = 2.5\times 10^{-5}$ 
rad$^2$/s$^3$.  This diffusivity comes from a fit of 
$\langle (d\dot\theta_0)^2\rangle = D_{\dot\theta}dt$ to experimental data for $\dot\theta_0(t)$ ($\langle (d\dot\theta_0)^2\rangle$ 
is the mean-square change in rotation rate over the time period $dt$). Again, this scaling only holds 
for intermediate time periods because $\dot\theta_0$ is bounded.

\begin{figure}
\includegraphics[width=2.5in]{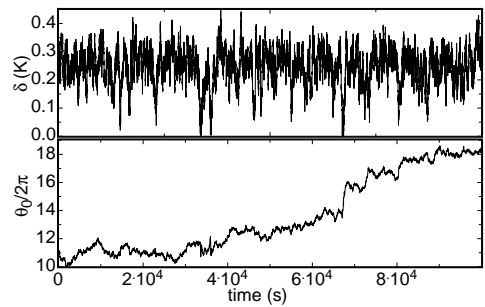}
\caption{A time series of $\delta$ and $\theta_0$ from the simulation of Eqs. \ref{eq:lang_delta} and 
\ref{eq:lang_theta}.  The LSC strength $\delta$ fluctuates around $\delta_0$, and occasionally those fluctuations are large enough to cause a cessation where $\delta \approx 0$. }
\label{fig:time_series}
\end{figure}

The two stochastic ODEs Eqs.~\ref{eq:lang_delta} and \ref{eq:lang_theta} are our model for the LSC dynamics. 
Using the experimentally determined values of $\delta_0, D_{\delta}$ and $D_{\dot\theta}$ and predictions 
for $\tau_{\delta}$ and $\tau_{\dot\theta}$ based on the measured value of $R_e$ discussed above, they can 
be integrated to get time series for 
$\delta$ and $\theta_0$. We used a simple first-order Euler method to solve the equations stepwise with a 
time step shorter than the smallest timescale $\tau_{\dot\theta}$ of the system.
Figure \ref{fig:time_series} shows a simulated time series over about one day.  One can see, as we expected, 
that the LSC amplitude $\delta$ is stable with an occasional cessation where the amplitude drops to zero.  
From much longer simulations we found that cessations occur about 3.8 times per day, which (we presume because the relation $\Delta V \gg D_{\delta}$ is not satisfied very well) is a factor of 3.7 less than the result from Eq.~\ref{eq:cess_rate} and about twice 
the frequency measured experimentally \cite{BA06}. Considering the approximations made in the model derivation, we regard this as very satisfying agreement with experiment.  

\begin{figure}
\includegraphics[width=2.5in]{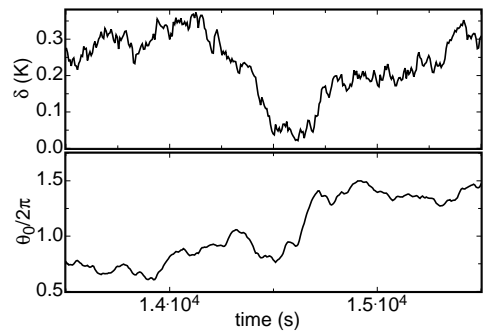}
\caption{A short section of the time series of $\delta$ and $\theta_0$ showing a cessation in detail.  The amplitude $\delta$ drops to near zero; while $\delta$ is small, the azimuthal motion becomes fast because it is no longer suppressed by the LSC angular momentum.}
\label{fig:cess_ex}
\end{figure}

Figure~\ref{fig:time_series} shows that the orientation 
meanders as expected, but one must look on a shorter time scale to see the details of the dynamics. Thus, Fig. \ref{fig:cess_ex} shows a shorter section of the same time series that contains a cessation.  One sees how $\delta$ 
gradually drops to zero, then grows back up again over a few hundred seconds, just as observed experimentally \cite{BA06}.  The time series for $\theta_0$ is interesting because there 
is a large change in $\theta_0$ during the cessation, again as seen experimentally \cite{BA06}.  Equation \ref{eq:lang_theta} for the azimuthal motion implies that, when  $\delta$ and thus the angular momentum are small during cessations, the damping term becomes small so the turbulent fluctuations are free to accelerate the LSC to large azimuthal rotation rates.  When the LSC is strong, i.~e.~$\delta \approx \delta_0$, then the larger angular momentum of the LSC supresses the azimuthal rotation.  This inverse relationship between $\delta$ and $\dot\theta_0$ was observed in experiments \cite{BNA05, BA06} but had not been explained by any previous model.

\begin{figure}
\includegraphics[width=2.25in]{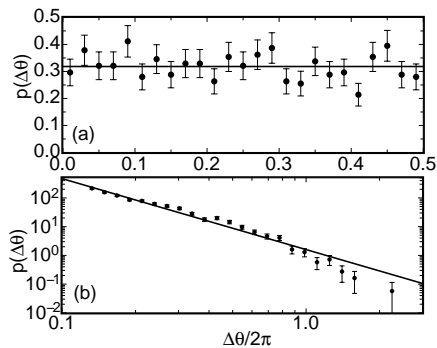}
\caption{(a)The probability distribution of the azimuthal change $\Delta\theta$ during cessations.  
The solid line represents the uniform distribution $p(\Delta\theta) = 1/\pi$. 
 (b) $p(\Delta\theta)$ for rotations.  The solid line is a power law.}
\label{fig:prob_dtheta}
\end{figure}

In order to determine the statistics of cessations and rotations, we analyzed the simulated time series 
using the same algorithms as those applied before to experimental data \cite{BNA05,BA06}. Figure 
\ref{fig:prob_dtheta}a shows the probability distribution $p(\Delta\theta)$ of the orientation change 
$\Delta \theta$ during cessations. 
The results from the simulations are consistent with a uniform distribution $p(\Delta\theta) = 1/\pi$, in 
agreement with the experiments \cite{BNA05, BA06}.  This is an important result that no model had predicted 
before, either because the $\theta$ dependence was not contained in it, \cite{SBN02,Be05} or because it was assumed that cessations would 
result in a reversal of the flow direction, i.e. $\Delta \theta = \pi$  \cite{FGL05}, or because the issue was not addressed \cite{RPTDGFL06}.  The angular momentum that usually 
suppresses the azimuthal motion of the LSC is reduced during cessations, allowing turbulent fluctuations to 
azimuthally rotate the LSC freely.  For large enough noise strengths, the azimuthal distance traveled is 
large over the duration of the cessation, and this results in a final orientation independent of the 
orientation before the cessation, which explains the uniform $p(\Delta\theta)$. 
We also find rotations in the model to result from a similar mechanism.  Rotations typically occur when 
the angular momentum of the LSC  is still large, so the azimuthal rotation in limited.  This results in a 
monotonically decreasing  distribution of $\Delta\theta$ for rotations.  However, the azimuthal rotation 
rate can become large even when $\delta$ is somewhat lower than normal, resulting in more rotations, and 
in particular more large rotations, than would be expected from purely Brownian noise with constant damping
coefficient \cite{BNA05}.  The 
simulation results can be represented reasonably well by a  power law for $p(\Delta\theta)$ as  shown in 
Fig~\ref{fig:prob_dtheta}b, in qualitative agreement with the experiments \cite{BNA05}.

To summarize, we derived a dynamical model motivated by the Navier-Stokes equations to describe the LSC in 
terms of two variables: its strength $\delta$ and its azimuthal orientation $\theta_0$.  The model used as 
input four parameters that were determined from independent experimental measurements, and required no adjustable parameters.  Each term in the equations has a clear physical meaning.  
The model produces a stable LSC with occasional cessations and rotations, and $p(\Delta\theta)$ is in 
agreement with experiments for both processes.  The frequency of cessations can be calculated based on an analogy to the 
Arrhenius-Kramers problem.  The azimuthal dynamics during cessations can be understood in terms of the angular
momentum of the LSC which suppresses azimuthal motion driven by turbulent fluctuations. The model also can 
reproduce other results for the LSC dynamics in good agreement 
with experiments; these are to be studied thoroughly in future work.

We benefited from conversations with numerous colleagues, especially with Detlef Lohse. This work was supported by the National Science Foundation through Grant DMR02-43336.


\begin{references}
\bibitem{FN_NL}In the absence of azimuthal motion the nonlinear term cancels in the volume average to be taken below. Its contribution may be neglected because the azimuthal motion is slow compared to $u_\phi$.
\end{references}
\end{document}